\documentstyle[editedvolume,psfig]{crckapbk}

\def \hi {H\,{\sc i~}}

\def \kms {km\,s$^{-1}$}
\def \NH {$N_{\rm HI}$}
\def \vlsr {$V_{\rm LSR}$}
\def \vgsr {$V_{\rm GSR}$}
\def \vlgsr {$V_{\rm LGSR}$}
\def \vdev {$V_{\rm DEV}$}

\def\deg{\hbox{$^\circ$}}
\def\arcmin{\hbox{$^\prime$}}

\def\fdg{\hbox{$.\!\!^\circ$}}
\def\farcm{\hbox{$.\mkern-4mu^\prime$}}

\begin{opening} \title{Compact, Isolated High--Velocity Clouds} \subtitle{}

\author{W.B. BURTON}
\institute{Leiden Observatory and
                 National Radio Astronomy Observatory}
\author{R. BRAUN}
\institute{Netherlands Foundation for Research in Astronomy}
\author{V. De\,Heij}
\institute{Leiden Observatory}

\runningtitle{Compact, Isolated High--Velocity Clouds}

\end{opening} 
\begin{document}

\begin{abstract} 
We consider here the class of compact, isolated, high--velocity \hi
clouds, CHVCs, which are sharply bounded in angular extent with no
kinematic or spatial connection to other \hi features down to a
limiting column density of 1.5$\times$10$^{18}$ cm$^{-2}$.  We describe
the automated search algorithm developed by de\,Heij et al.  (2002a)
and applied by them to the Leiden/Dwingeloo Survey north of $\delta=
-28^\circ$ and by Putman et al. (2002) to the Parkes HIPASS data south
of $\delta=0\deg$, resulting in an all--sky catalog numbering 246
CHVCs.  We argue that these objects are more likely to represent a
single phenomenon in a similar evolutionary state than would a sample
which included any of the major HVC complexes.  Five principal
observables are defined for the CHVC population: \,(1) the spatial
deployment of the objects on the sky, (2) the kinematic distribution,
(3) the number distribution of observed \hi column densities, (4) the
number distribution of angular sizes, and (5) the number distribution
of line widths.  We show that the spatial and kinematic deployments of
the ensemble of CHVCs contain various clues regarding their
characteristic distance.  These clues are not compatible with a
location of the ensemble within the Galaxy proper. The deployments
resemble in several regards those of the Local Group galaxies.

We describe a model testing the hypothesis that the CHVCs are a Local
Group population. The agreement of the model with the data is judged by
extracting the observables from simulations, in a manner consistent
with the sensitivities of the observations and explicitly taking
account of Galactic obscuration.  We show that models in which the
CHVCs are the \hi counterparts of dark--matter halos evolving in the
Local Group potential provide a good match to the observables, if
account is taken of tidal and ram--pressure disruption, the
consequences of obscuration due to Galactic \hi and of differing
sensitivities and selection effects pertaining to the surveys.

A representative sample of CHVCs has been studied with high angular
resolution (sub-arcminute) using the WSRT and with high \NH~sensitivity
($<10^{17}$\,cm$^{-2}$) using the Arecibo telescope. The picture that
emerges is a nested morphology of CNM cores shielded by WNM cocoons,
and plausibly surrounded by a Warm Ionized Medium halo.  These
observations lead to indirect contraints on the distances, ranging from
150 to 850 kpc.
\end{abstract}

\section{Introduction}

\hi structures found lying in the velocity regime harboring the
high--velocity clouds manifest themselves in different forms;
considered collectively, these clouds either do not represent a single
astronomical phenomenon, or (more likely) do not represent the
phenomenon at a single stage in its evolution, with all members seen
under the same physical circumstances.  The variety of positional and
kinematic properties may correspond to differing evolutionary
histories.  The Magellanic Stream, for example, is identified as
associated with the Large Magellanic Cloud because of positional and
kinematic coincidences of part of the Stream with the LMC.  It is
composed of material either tidally stripped from the Milky Way or from
the LMC itself, or gravitationally captured from elsewhere, but in any
case constrained to follow the orbit of the LMC, and thus presumably
lies at the LMC distance of some 50 kpc.  There are several other major
streams of HVCs, notably Complexes M, C, and A, which can also be
traced over many tens of degrees and which, like the Magellanic Stream,
individually show substantial spatial and kinematic coherence.  The
distance of only one complex has been bracketed with any certainty:
absorption--line measurements (van Woerden et al. 1999, Wakker 2001)
toward Complex A place it within the range $4 < d < 10$ kpc.  The
histories of the Magellanic Stream and of Complex A (and probably of
the other large complexes as well) have been multifaceted, with
exposure to strong radiation fields and tidal distortions by the Milky
Way.  Braun \& Burton (1999; BB99) identified CHVCs as a subset of the
anomalous--velocity gas which might be characteristic of a single class
of objects, whose members plausibly originated under common
circumstances and share a common subsequent evolutionary history.  The
discussion of the CHVCs hypothesized that the large HVC complexes
(other than the Magellanic Stream) were once similar objects, but which
now are relatively nearby.

We describe in \S\,2 the search algorithm that has been applied to
the all--sky coverage afforded by recent \hi surveys of the northern and
southern skys.  In \S\,3 we show the principal observable quantities of
the all--sky ensemble, and briefly remark on conclusions which follow
directly from these observables.  In \S\,4, we show that a model in
which the CHVCs are the \hi counterparts of dark--matter halos evolving
in the Local Group potential provides a good match to the observables,
but only if the simulation is sampled as if it were being observed with the
sensitivities of the available observations and if explicit account is
taken of obscuration by \hi in the Galaxy. In \S\,5, we show WSRT and
Arecibo high--resolution imaging of a selection of individual CHVCs,
stressing the importance of the synthesis observations to study of the cold
cores and of the single--dish observations to study of the diffuse halos.
The high--resolution imaging supports several indications of substantial
distances.

\section{An all-sky CHVC catalog extracted from the LDS and
HIPASS}

Obtaining an adequate observational foundation for the
anomalous--velocity \hi clouds has been a persistent and continuing
challenge.  The BB99 sample was obtained by visual inspection of the
Leiden/Dwingeloo Survey (LDS, Hartmann \& Burton 1997).  The LDS was
observed with the 25--m Dwingeloo telescope, whose {\sc fwhm} beam
subtends $36\arcmin$, covering the sky north of $\delta = -30\deg$ on a
$0\fdg5$ grid.  The effective velocity coverage of the LDS spans $-450
< $ \vlsr~$ < +400$ \kms, resolved into spectral channels of 1.03
\kms~width.  The nominal brightness--temperature sensitivity is 0.07 K.
 
The Parkes All--Sky Survey (HIPASS, Barnes et~al. 2001) advanced the
observational foundations of HVC work.  The HIPASS material was Nyquist
sampled at the angular resolution of~$15\farcm5$ afforded by the Parkes
64--m telescope, covering the sky south of $\delta=0\deg$ and reaching
a $5\sigma$~{\sc rms} brightness--temperature sensitivity of
approximately 50~mK.  The spectra cover the range $-700 < $ \vlsr~$ <
+500$ \kms, with channels separated by 26 \kms.  The sensitivity,
angular resolution, and extent of velocity coverage of the HIPASS
material are superior to those of the LDS, but the velocity resolution
and ability to recovery the most extended spatial structures of the LDS
is superior to that of the HIPASS. We comment below on the
inhomogeneities introduced in our all--sky CHVC catalog by the
differing observational parameters of the two surveys.

An automated algorithm for extracting anomalous--velocity \hi clouds
from survey material has been developed by de\,Heij et al. (2002a), and
applied by them to the LDS and by Putman et al. (2002) to the HIPASS.
The algorithm sought objects isolated in position and in velocity, down
to \NH~values below about $1.5\times10^{18}$cm$^{-2}$.  Initial
application of the algorithm resulted in a list of all objects
satisfying the search criteria, and thus included not only compact
high--velocity clouds, but also structures that are part of HVC
complexes, some intermediate--velocity features, and even the gaseous
disk of our Galaxy, as well as features which were subsequently
eliminated as due to excessive noise, radio interference, the
non--square response of the receiver bandpass, or other imperfections
in the data.

To avoid possible contamination from emission associated with our
Galaxy and the intermediate--velocity complexes, all clouds which did
not satisfy a deviation--velocity constraint were removed from the
preliminary catalog.  Wakker (1990) defined \vdev~as the excess
velocity of a feature compared to velocities allowed by a simple model
of our Galaxy.  In order to constrain possible Milky Way contamination
more accurately than would be possible with a constant value of \vdev,
we modified the definition, allowing the limiting velocity to vary with
direction, in order to account for the warp of the Milky Way \hi disk,
for its flare to increasing thickness with increasing Galactocentric
distance, and for its lopsidedness in plan view.  The gas--layer model
was derived from the Voskes \& Burton (1999) analysis of combined
southern and northern \hi survey data. De\,Heij et al. (2002b) describe
how discrimination against CHVCs in the ``\hi~zone of avoidance"
(which, unlike the optical counterpart, is velocity--dependent as well
as position--dependent) influences the subsequent analysis: the swath
of this zone of avoidance is distorted after coordinate transformation
from the $V_{\rm LSR}$ to the \vgsr~or \vlgsr~frames, corresponding to
the Galactic Standard of Rest or the Local Group Standard of Rest,
respectively, which are shown by the kinematics to be more relevant to
the CHVC ensemble.

The degree of isolation of the clouds extracted using the algorithm was
determined from velocity--integrated images covering an area measuring
$10\deg$ by $10\deg$ centered on the position of the cloud in question
and spanning a matched velocity interval. An ellipse was fit to the
contour with a value of half the maximum brightness of the cloud to
allow tabulation of a characteristic size and orientation.  The degree
of isolation was assessed on the basis of the lowest significant
contour level of \NH~ commensurate with the data sensitivity,
corresponding to about $1.5\times10^{18}$cm$^{-2}$ over the typical
25~\kms FWHM linewidth in both the LDS and HIPASS surveys.  We demanded
that this contour be closed, with its greatest radial extent less than
the $10\deg$ by $10\deg$ image size, and that there not be confusion
from any nearby extended emission near our cut-off column density.
Both the degree of isolation and the total angular size (measured at
our cut-off level) will therefore be dependent on data
sensitivity. While unfortunate, this is an inevitable limitation of
real data. A distinction was made between the cleanest isolated
objects, CHVCs, and the somewhat ambiguous candidates, CHVC:s and
CHVC?s, which had some degree of confusing emission in their $10\deg$
by $10\deg$ environment.  While isolated objects as large as about
8$^\circ$ could have been found by our algorithm, the distribution is
strongly peaked with a median angular size of only $1\deg$ and a
maximum FWHM of 2.$^\circ$2. This small median angular size motivates
the use of the term ``compact'' in the object designation.  It is not
ruled out that some of the features which we tabulated as isolated in
$l,b,V$ space would be shown under scrutiny of more sensitive data to
be embedded in a weaker envelope, or even to be part of a large, but
relatively weak, complex or stream.  And it is indeed expected that
deeper observations will result in detection of additional CHVCs: the
modelling discussed by de\,Heij et al. (2002b), and briefly summarized
below, leads to specific predictions in this regard.

Independent confirmation was required for all of the CHVCs in the LDS
portion of the all--sky catalog.  The LDS is not free of a pernicious
type of radio interference, which occasionally produced a broad, weak
spectral feature, of the same general form as the signature of a
high--velocity cloud, or of a nearby dwarf galaxy.  This RFI was
sometimes of short temporal duration, not long enough that a persistent
telltale spoor in a single frequency band would reveal the signal as
spurious.  Confirmation is all the more important because the isolated
objects being sought appear commonly at only a few LDS lattice points,
or even at only one.  Confirmation was not required in the HIPASS case,
because that survey accumulated independent data in the multi--beam
feeds. All of the LDS CHVCs were confirmed either from published data
or from new observations made with the WSRT, operating in a
total--power observing mode whereby auto--correlation spectra were
obtained from all 14 individual 25--meter antennas, rather than the
more usual cross-correlation spectra.
 
The all--sky catalog of CHVCs includes 67 found in the LDS at
$\delta>0\deg$ by de\,Heij et al. (2002a) and 179 found in the HIPASS
material at $\delta<0\deg$ by Putman et al. (2002). The data on CHVCs
found by both surveys in the overlap zone $-30\deg <\delta <0\deg$ are
recorded in the all-sky catalog using the HIPASS parameters. The
detection statistics in the overlap zone supports discussion of the
relative completeness of the two surveys.

\section{Principal morphological observables of the ensemble of CHVCs}

\subsection{Observed spatial deployment of the CHVCs }

\begin{figure}
\caption{Spatial deployment of CHVCs over the sky.  Triangles represent
the LDS sample cataloged by de\,Heij et al. (2002) at $\delta>0\deg$;
diamonds represent the HIPASS sample of Putman et al.  (2002a) at
$\delta<0\deg$. Filled circles correspond to the galaxies comprising
the Local Group, as tabulated by Mateo (1998).  Red symbols indicate
\vlsr\,$>0$\,\kms; black ones, \vlsr\,$<0$\,\kms. The background greys
show the \hi column depths from an integration of observed temperatures
over velocities ranging from $V_{\rm LSR} = -450$\,\kms~to
$+400$\,\kms, but excluding emission with $V_{\rm DEV} < 70$\,\kms.
The larger number of southern CHVCs, and specifically the apparent
demarcation in the source densities at $\delta = 0\deg$, are largely
due to the differing responses of the two data sets, as discussed
briefly in the text and reproduced by our modelling.  }
\end{figure}

Figure 1 shows the distribution on the sky of the CHVCs entered in the
combined LDS and HIPASS catalog.  The CHVCs show no obvious tendency to
cluster in streams or complexes. In particular, there is no preference
in the sky distribution for accumulation near the Galactic equator, or
towards the inner Galaxy, or in association with any of the known HVC
complexes or with the Magellanic Stream. Unlike the situation
pertaining for the principal HVC complexes, which show a substantially
higher total \hi flux in the northern Galactic hemisphere than in the
southern, the CHVCs in the all--sky catalog are more numerous in the
southern hemisphere than in the northern.  The over--density in the
southern hemisphere is due principally to the higher sensitivity of
the HIPASS material compared to the LDS; but, as we mention briefly
below and discuss in more detail in de\,Heij et al. (2002b), a slight
enhancement remains after taking the observational parameters into
account.  The locations of Local Group galaxies as tabulated by Mateo
(1998) are plotted in Fig. 1 as filled circles.  Although the
statistics are weak, to first order the sky deployment of CHVCs
resembles that of the galaxies.

\subsection{Observed kinematic deployment of the CHVCs  }

\begin{figure}
\caption{Kinematic deployment and perceived size-- and column--depth
distributions of the CHVC ensemble.  These properties, together with
the spatial deployment plotted in Fig. 1, constitute the principal
observables of the ensemble. The upper two panels show the variation of
\vgsr~plotted against Galactic longitude, and against latitude,
respectively. Triangles represent CHVCs identified at $\delta>0\deg$ in
the LDS, and diamonds represent CHVCs found in the HIPASS; filled
circles show the kinematic distribution of the Local Group galaxies
listed by Mateo (1998). The histograms in the bottom row show the
observed angular size distribution of the CHVC population, on the left,
and the observed peak \hi column--density distribution, on the right. }
\end{figure}

The longitude distribution of the CHVC motions, measured with respect
to the GSR frame, is shown in the upper panel of Fig. 2; the latitude
distribution of \vgsr~is shown in the middle panel.  Also plotted is
the velocity/longitude distribution of the galaxies in the Local Group.
The CHVCs are confined within a definite kinematic envelope, narrower
in extent than the spectral coverage of the surveys; we stress below
that this confinement is not a selection effect.

The CHVC velocities can be considered in different reference frames.
The mean velocity for the all--sky ensemble of CHVCs measured in the
LSR frame is $-45$\,\kms; although not clearly centrally peaked, the
dispersion of the \vlsr~kinematics is formally 238\,\kms.  In the GSR
frame, the mean velocity decreases to $-63$\,\kms, with a dispersion of
128\,\kms; in the Local Group Standard of Rest frame, the mean velocity
is $-60$\,\kms, with a dispersion of 112\,\kms.  The mean velocities of
the 27 dwarf galaxies in the Local Group with known radial velocities
tabulated by Mateo (1998) are $-57$, $-22$, and $+4$\,\kms, in the LSR,
GSR, and LGSR reference frames, respectively, with the corresponding
values of the dispersion being 196, 104, and 78\,\kms. The net infall
of the CHVC ensemble seems firmly established.  Although the dispersion
of CHVC velocities is formally somewhat less in the LGSR frame than in
the GSR one, it will require more sensitive data before the
significance of this difference can be judged.

Taken at face value, these kinematics are suggestive of a population
which is bound to the Local Group, but which has not yet experienced
significant interaction or merger with the larger members, which would
virialize their motions. A mean radial infall velocity which is
comparable to the velocity dispersion of the Local Group member
galaxies, is plausible given that the same gravitational potential is
responsible for both.  Perhaps of relevance is the observation by
C\^ot\'e et al. (1997) that the faint dwarf galaxies in the two groups
of galaxies nearest to the Local Group show a wider range of velocity
distribution than the brighter members. An evolutionary history which
is a function of mass may be a natural part of the galaxy formation
process.

\subsection{Observed size and column density distributions}

In addition to the spatial and kinematic distributions, the measured
CHVC sizes, line widths, and column depths are also important
observables against which interpretations must be judged.  The
distributions of CHVC sizes and column depths are shown in the lower
panels of Fig. 2. Although the only limit on angular size we have
imposed is the $10\deg \times 10\deg$ dimension of our initial
\NH~image of each candidate, the size distribution is nevertheless
strongly peaked with a median at $1\deg$ {\sc fwhm}, and furthermore
does not extend beyond $2\fdg2$.  Sharply bounded anomalous--velocity
\hi objects apparently are, indeed, compact.  The distribution of
velocity {\sc fwhm} is strongly peaked at about 25 \kms, although the
distribution does extend out to 100~\kms.  Most CHVCs are spatially
unresolved in the LDS beam, and many are unresolved in the HIPASS beam;
although only some 20 CHVCs have been imaged at higher angular
resolution, the indications are that the broadest widths are
contributed by differing systemic velocities of several individual
cores, each with a relatively narrow width, constituting a single CHVC.

\subsection{Comments on sample completeness and homogeneity}
 
Our CHVC tabulation is probably not incomplete as a consequence of the
velocity range limits of the observations.  Although the LDS was
searched only over the range $-450 < V_{\rm LSR} < +350$ \kms, there
are indications that we did not miss many (if any) clouds because of
this limited interval.  The feature with the most extreme negative
velocity yet found is CHVC\,$110.6\!-\!07.0\!-\!466$, discovered by
Hulsbosch (1978); the Wakker \& van Woerden (1991) tabulation, which
relied on survey data covering the range $-900 < V_{\rm LSR} < +750$
\kms, found no high--velocity cloud at a more negative velocity.  The
HIPASS search by Putman et al. (2002) sought anomalous--velocity
emission over the range $-700 < V_{\rm LSR} < +1000$ \kms.  Ten of the
CHVCs cataloged by Putman et al.  have $V_{\rm LSR}<-300$ \kms, but the
most extreme negative velocity is only $-353$ \kms, for
CHVC\,$125.1\!-\!66.4\!-\!353$.

In terms of the Local Group deployment variously discussed by Blitz et
al.  (1999), BB99, Putman \& Moore (2002), and de\,Heij et al. (2002b),
the most extreme negative velocities would be found in the general
region of the barycenter of the Local Group, whereas the most extreme
positive LSR velocities, which would also be more modest in amplitude
than the extreme negative velocities, would be expected in the general
region of the Local Group anti--barycenter.  Only 7 of the HIPASS CHVCs
have $V_{\rm LSR}$ greater than $+300$ \kms, and only one has a
velocity greater than $350$ \kms, namely
CHVC\,$258.2\!-\!23.9\!+\!359$.  All of the 7 CHVCs with substantial
positive velocities lie deep in the third longitude quadrant, or in the
early fourth, and thus in the general vicinity of the anti--barycenter.

In view of these detection statistics, it seems unlikely that the
velocity--range limits of either the LDS or of the HIPASS have caused a
significant number of CHVCs to be missed.  Thus, the true velocity
extent, as well as the non--zero mean in the LSR frame, of the
anomalous--velocity ensemble appear well represented by the extrema of
$-466$ \kms~and $+359$ \kms.

It is appropriate to comment on certain selection effects which might
cause systematic distortions to the true distribution of CHVCs.  A
demarkation at $\delta = 0^\circ$ separates the CHVC catalog entries
based on the LDS from those based on HIPASS.  The weaker sensitivity of
the LDS is likely to have discriminated against some CHVCs, in
particular those at positive velocities: the color--coding of CHVC
velocities shown in Fig. 1 indicates that more positive--velocity
objects are expected in Galactic quadrant II than in the other
quadrants.  Quadrant II is almost exclusively LDS's territory, not
HIPASS's, and is therefore prejudiced by the weaker sensitivity of the
LDS.  There is also a mild asymmetric discrimination against detecting
CHVCs due to their submersion in the ``\hi zone of avoidance",
i.e. near $V_{\rm LSR} = 0$\,km/s, with the velocity discrimination
slightly skewed to positive $V_{\rm GSR}$--values in the GSR frame, as
discussed by de\,Heij et al. (2002b).

If the HIPASS and LDS CHVC (or HVC) catalogs are compared or if they
are used together, for example to investigate the all--sky properties
of the CHVC ensemble, then due attention should be given to the higher
expected detection rate in the southern material, because of its
greater sensitivity and denser angular sampling.  To correct for the
higher detection rate in the south, de\,Heij et al. (2002b) considered
the likelihood that a given cloud which is observed in the HIPASS data
also will be observed in the LDS. This issue could be directly
confronted, since the two surveys overlap in the declination range
$-28\deg \leq \delta \leq +2\deg$.  Comparison of the results derived
by applying the search algorithm to both surveys allows assessment of
the robustness of the selection criteria and of the completeness of the
LDS catalog.

The HIPASS {\sc rms} noise figure is 10~mK, for a channel 26 \kms~wide
and a {\sc fwhp} beam of $15^\prime$; the corresponding LDS {\sc rms}
value is 70~mK, for a channel width of 1.03 \kms~and a {\sc fwhp} beam
of $36^\prime$.  After smoothing both surveys to the same spectral
resolution of 26 \kms, the 3$\sigma$ limiting column density (for
emission filling each beam) is 0.47 and 0.64$\times$10$^{18}$ cm$^{-2}$
in the HIPASS and LDS, respectively. Thus, while the sensitivity to
well--resolved sources is comparable, the point--source sensitivity of
HIPASS is greater by about a factor of 3. On the other hand, the
band--pass calibration of the HIPASS data relies on reference spectra,
deemed empty of \hi emission, that are offset by only a few degrees on
the sky. Even with the {\sc minmed5} method of baseline determination
employed by Putman et al., there is significant filtering of extended
emission, which complicates the assessment of object isolation down to
a low column density limit.

The velocity resolution of the Leiden/Dwingeloo survey is 1.03 \kms,
and this resolution should suffice, as all known CHVCs have broader
velocity widths.  Although the {\sc fwhm} of most of the CHVCs
tabulated here is greater than 20 km/s, a few are considerable
narrower, and it is not unexpected that some would be missing from the
HIPASS compilation because they are diluted by the relatively coarse
HIPASS channel spacing.

The differences in point--source sensitivity on the one hand and
sensitivity to very extended structures on the other, results in a
larger number of faint--source detections in HIPASS, but also in a
different designation for some of the brighter clouds which the surveys
have in common. Comparison of the results derived by applying the
search algorithm to both surveys allows assessment of the robustness of
the selection criteria and of the completeness of the LDS catalog.  The
comparison discussed by de\,Heij et al. (2002b) shows the influence of
survey sensitivity on the number of objects found.  Above a peak
temperature of~0.45~K, the LDS results are as complete as those based
on HIPASS; between $T_{\rm peak}=0.20$ K and 0.45 K, the LDS results
recovered 83\% of the clouds found in the HIPASS data.  The
completeness of the LDS catalog drops rapidly at lower values of the
peak temperature: HIPASS clouds less bright than 0.20 K are almost
completely absent from the LDS catalog.  The incompleteness at low peak
temperatures will be more important for very compact CHVCs. The
temperatures and therefore sensitivities were measured at the grid
points of the survey; point sources that are not located at a grid
point will have been observed with a reduced sensitivity, which depends
on the telescope beam and the distance to the nearest grid point.  The
LDS sampled the sky on a $0\fdg5$ lattice; the sensitivity away from
the grid points falls off as a Gaussian with a {\sc fwhp}
of~36$^\prime$. The small sizes of the CHVCs and the
less--than--Nyquist sampling interval of the LDS can conspire to result
in an observation with a lower signal--to--noise ratio in the LDS than
in the HIPASS.

\subsection{Initial conclusions from the ensemble observables}

Some general conclusions may be drawn from an initial inspection of the
all--sky CHVC catalog.  We noted above the roughly even distribution
over the sky, the constrained kinematics within a definite envelope,
the net infall towards the Galaxy, and the substantially smaller velocity
dispersion in the GSR and LGSR frames than in the LSR one.

The characteristic \NH~shown in Fig. 2 is also of interest.  The
significance of \hi features which are isolated in column density down
to a level as low as $1.5\times$10$^{18}$ cm$^{-2}$, is that this is
about an order of magnitude lower than the critical column density
identified at the edges of nearby galaxies (Maloney 1993, Corbelli \&
Salpeter 1993), $\sim$2$\times$10$^{19}$ cm$^{-2}$, where the ionized
fraction is thought to increase dramatically due to the extragalactic
radiation field.  These isolated objects will need to provide their own
shielding to ionizing radiation.  In this case, their small median
angular size, of about $1^\circ$ {\sc fwhm}, is consistent with
substantial distances, since the partially ionized \hi skin in a
power--law ionizing photon field has a typical exponential
scale--length of 1~kpc (Corbelli \& Salpeter 1993).

The entries in the CHVC catalog show spectral {\sc fwhm} values ranging
from 5.9 km/s, characteristic of a very narrow, cold \hi feature in the
conventional gaseous disk, to 95.4 km/s, characteristic of some
external galaxies. Interferometric observations of some of the broad
objects have shown that the large width is contributed by individual
cores within a common envelope.  But the distinction between the \hi
properties of CHVCs and dwarf galaxies with very weak star formation,
or proto--galaxies with no star formation, remains to be made.  The
broadest object found in the BB99 search for CHVCs was revealed as a
low--surface--brightness spiral galaxy (Burton et al. 1998).  Warned by
the presence this interloper, we searched the Digital Sky Survey in the
direction of each of the CHVCs listed in the LDS portion of the
catalog, but found no optical counterparts to the objects listed; a
recent deeper search by Simon \& Blitz (2002) also found no optical
counterparts.  The narrowest widths, even without higher--resolution
imaging, serve to meaningfully constrain the kinetic temperature of the
gas, as well as the line--of--sight component of any rotation or shear
in a single object, and the range of kinematics if the object should be
an unresolved collection of subunits moving at different velocities.

CHVCs near the Galactic equator display the horizontal component of
their space motion. The middle panel of Fig. 2 shows that the radial
motions at low $|b|$ are as large as those at high latitudes, and
furthermore that the CHVC distribution does not avoid the Galactic
equator. Large horizontal motions are difficult to account for in terms
of a Galactic fountain (Shapiro \& Field 1976, Bregman 1980).  Burton
(1997) has noted that anomalous--velocity clouds do not contaminate the
\hi terminal--velocity locus in ways which would be expected if they
pervaded the Galactic disk, and that this observation constrains the
clouds either to be an uncommon component of the Milky Way disk,
confined to the immediate vicinity of the Sun, or else to be typically
at large distances beyond the Milky Way disk.  We note also that the
lines of sight in the directions of each of the low--$|b|$ CHVCs
traverse some tens of kpc of the disk before exiting the Milky Way:
unless one is prepared to accept these CHVCs as boring through the
conventional disk at hypersonic speeds (for which there is no
evidence), and atypical in view of the cleanliness of the
terminal--velocity locus, then their distances are constrained to be
large.

Similarly, CHVCs located near the Galactic poles offer unambiguous
information on the vertical, $z$, component of their space motion. The
vertical motions are substantial, with recession velocities
approximately equal in number and amplitude to approach velocities; the
vertical motions are, furthermore, of approximately the same amplitude
as the horizontal ones.  This situation also is incompatible with the
precepts of the fountain model, which predicts negative
$z$--velocities, for material returning in a fountain
flow. Furthermore, the values of the $z$--velocities are predicted not
to exceed the velocity of free fall, of some 200 \kms. In fact,
vertical motions of twice the free fall value are observed.  We note
that some of the CHVC objects are even moving with velocities in excess
of a plausible value of the Milky Way escape velocity (cf. Oort 1926).

The aspects of the spatial and kinematic topology of the class
mentioned above are difficult to account for if the CHVCs are viewed as
a Milky Way population, in particular if they are viewed as
consequences of a Galactic fountain (see also Blitz 2001). These same
aspects would seem to discourage revival of several of the earlier
Milky Way mechanisms (reviewed, for example, by Oort 1966) suggested
for the general HVC phenomenon, including ejection from the Galactic
nucleus, association with a Galactic spiral arm at high latitude, and
ejection following a nearby supernova explosion.

\section{The CHVC ensemble modeled as a Local Group population, with the
model sampled as if observed}
 
The spatial and kinematic distributions of the all--sky CHVC ensemble
are consistent with a dynamically cold population spread throughout the
Local Group, but with a net negative velocity with respect to the mean
of the Local Group galaxies.  BB99 suggested that the CHVCs represent
the dark matter halos predicted by Klypin et al.  (1999) and Moore et
al. (1999) in the context of the hierarchical structure paradigm of
galactic evolution.  These halos would contain no stars, or only a few;
most of their visible matter would be in the form of atomic hydrogen.
Although many of the halos would already have been accreted by the
Galaxy or M\,31, some would still populate the Local Group, either
located in the far field or concentrated around the two dominate Local
Group galaxies.  Those passing close to either the Milky Way or M\,31
would be ram--pressure stripped of their gas and tidally disrupted by
the gravitational field. Near the Milky Way, the tidally distorted
features would correspond to the HVC complexes observed.

\subsection{Description of the Local Group model}

\begin{figure}
\caption{ {\it left panel:} Average velocity field in the Local Group
model of CHVC deploment described in the text. The Milky Way is located
at $x=-0.47$ Mpc, $y=0.0$ Mpc, and M\,31 at $x=0.23$ Mpc, $y=0.0$ Mpc.
The distribution of test particles representing the CHVCs is given by
two Gaussian functions, one centered on the Milky Way and the other on
M\,31; the contours show the space density of the CHVC test particles.
The vectors give the average velocity of test particles in a $10 \times
10$\,kpc box centered on each gridpoint. (The length of the thick--line
vector in the upper left corner corresponds to 200 \kms.)  Squares are
drawn if the velocity dispersion of the particles within the box
exceeds 100\,\kms.  {\it right panel:} Distribution of synthetic CHVCs
shown in an $(x,y)$ projection, for a realization of a Local Group
simulation corresponding to the parameters described in the text. The
Galaxy is located at the origin; crosshairs locate M\,31.  A symbol
marks the location of each of the input CHVCs entering the simulation.
But not all of the input clouds survive the simulated environment:
black symbols indicate those input clouds which are destroyed by tidal
and ram--pressure stripping influences of M\,31 and the
Galaxy. Furthermore, not all surviving input clouds are sufficiently
intense to be detected by surveys with the properties of the LDS or
HIPASS: grey symbols indicate surviving input clouds which would be too
weak to be detected by the LDS or by HIPASS, respectively, depending on
their declination. Red symbols indicate those input clouds which would
be perceived as CHVCs if the simulation were sampled as if by the LDS
and HIPASS.}
\end{figure}

Recently several simulations have been performed to test the hypothesis
that the CHVCs are the remaining building blocks of the Local Group.
Putman \& Moore (2002) compared the results of the full n--body
simulation described by Moore et al. (2001) with various spatial and
kinematic properties of the CHVC distribution, as well as with
properties of the more general HVC phenomenon, without regard to object
size and degree of isolation.  Putman \& Moore were led to reject the
Local Group deployment of CHVCs, for reasons which we debate
below. Blitz et al. (1999) had earlier performed a restricted two--body
analysis of the motion of clouds in the Local Group.  In their attempt
to model the HVC distributions, Blitz et al.  modeled the dynamics of
dark matter halos in the Local Group and found support for the Local
Group hypothesis when compared qualitatively with the deployment of a
sample of anomalous--velocity \hi containing most HVCs, but excluding
the large complexes and the Magellanic Stream, for which plausible or
measured distance constraints are available.

De\,Heij et al. (2002b) followed the modeling approach of Blitz et al.,
but judge the results of their simulations against the properties of
the all--sky CHVC sample, viewing the simulated data {\it as if it were
observed} with the LDS and HIPASS surveys. A simulation was run for
each set of model parameters. We chose a position for each object, in
agreement with the spatial density distribution of the ensemble, but
otherwise randomly. The velocity was given by the velocity field
illustrated in Fig. 3.  The \hi mass of a test cloud was randomly set
in agreement with a power--law mass distribution between a specified
upper mass limit and a lower mass limit. The physical size and line
width of each object follow from the choice of the power--law index,
$\beta$. Once all these parameters were set, we determined the observed
peak column density and angular size. Objects in the northern
hemisphere were convolved with a beam appropriate to the LDS, while
those at $\delta < 0\deg$ were convolved with the HIPASS beam.
Simulated objects were considered detected if the peak brightness
temperature exceeded the detection threshold of the relevant survey,
i.e. the LDS for objects at $\delta >0\deg$, and HIPASS otherwise.
Furthermore, in order for a test object to be retained as detected its
deviation velocity was required to exceed $70\rm\;km\;s^{-1}$ in the
LSR frame. In addition, each simulated cloud was judged whether or not
it was stable against both tidal disruption and ram--pressure stripping
by the Milky Way and M\,31; if not, it was removed from further
consideration.  A cloud was regarded stable against ram--pressure
stripping if the gas pressure at its center exceeded the ram pressure,
$P_{\rm ram} = n_{\rm halo} \cdot V^2$, for a cloud moving with
velocity~$V$ through a gaseous halo with density $n_{\rm halo}$.  A
cloud was considered tidally disrupted if the gravitational tidal field
of either the Galaxy or M~31 exceeded the self--gravity of the cloud at
its location. We continued simulating additional objects following this
prescription until the number of detected model clouds was equal to the
number of CHVCs in our observed all--sky sample.

The best--fitting populations found by de\,Heij et al. have a maximum
HI mass of $10^7\rm\;M_\odot$, a slope of the HI mass distribution in
the range $-1.7$ to~$-1.8$, and a Gaussian dispersion for their spatial
distributions of between 150 and 200~kpc centered on both the Milky Way
and M\,31. Given its greater mean distance, only a small fraction of
the M\,31 sub--population is predicted to have been detected in present
surveys.

\begin{figure}
\caption{ Sky deployment of synthetic CHVCs corresponding to the
simulation of a population in the Local Group with the parameters
described in the text.  Black dots correspond to input clouds which
would not enter a catalog of CHVCs based on the LDS or HIPASS data,
either because the input clouds have been destroyed (tidally, or by
ram--pressure stripping) as a consequence of their proximity to M\,31
or to the Galaxy, or else because their perceived flux would be below
the detection thresholds of the LDS or HIPASS, depending on their
declination.  Red symbols indicate clouds which would be detected were
the simulation sampled as if by the LDS or HIPASS.  The background grey
image, drawn as in Fig. 2, shows the demarcation between the lower
sensitivity LDS regime and the higher sensitivity HIPASS regime.  Many
of the input clouds are clustered around M\,31 and the region of the
barycenter of the Local Group, and lie thus in the northern regime of
the sky observed by the LDS where the simulated observations are
relatively less sensitive.  The magnitude of the over--density of red
dots in the southern hemisphere does not reflect the true distribution
of input CHVCs, but the combined effects of the simulated physical
environment and the observational parameters of the LDS and HIPASS
surveys. The sky deployment of the simulated CHVC population can be
compared with the observed situation shown in Fig. 1.}
\end{figure}

\subsection{Simulated CHVC distributions, perceived as if observed}

\begin{figure}
\caption{ Kinematic deployment and perceived size-- and column--depth
distributions of the simulated CHVC ensemble in the Local Group,
plotted for comparison with the observed distributions shown in Fig. 2.
The two upper panels show the variation of $V_{\rm LSR}$ with Galactic
longitude and latitude, respectively.  Black symbols refer to input
clouds which do not survive in the Local Group environment or which do
not have sufficient flux to be detected by the Leiden/Dwingeloo or
Parkes surveys; red symbols indicate clouds which would be detected if
the input ensemble were sampled with the observational parameters
characterizing the LDS and HIPASS data.  The two panels in the bottom
row show the perceived angular size distribution of the simulated CHVC
population, on the left, and the peak \NH~distribution which is
perceived in the simulation, on the right. }
\end{figure}

An appreciation of the physical appearance of these Local Group models
is provided by the righthand panel of Fig. 3, showing a perpendicular
projection of the model population. The $(x,y)$ plane in the figure is
the extended Galactic plane, with the Galaxy centered at $(x,y)=(0,0)$
with the positive $z$ axis corresponding to positive $b$. The intrinsic
distribution of objects is an elongated cloud encompassing both the
Galaxy and M\,31, dominated in number by the M\,31 concentration. The
objects that have at some point in their history approached so closely
to either of these galaxies that their \hi would not survive the
ram--pressure or tidal stripping are indicated by the filled black
circles. Cloud disruption was substantially more important in the M\,31
concentration than in the Galaxy concentration. The objects that are
too faint to have been detected by the LDS or HIPASS observations,
depending on declination, are indicated by grey circles. {\emph {The
bulk of the M\,31 sub--concentration is not detected in our CHVC
simulation for two reasons}}:~\,(1) these objects have a larger average
distance than the objects in the Galactic sub--concentration, and (2)
the M\,31 sub--concentration is located primarily in the northern
celestial hemisphere, where the lower LDS sensitivity compromises
detection.  The objections posed by Putman \& Moore (2002) to the Local
Group deployment models, based on the failure to observe a clustering
of CHVCs near M\,31 or the Local Group barycenter, seem accounted for
by the crucial role of survey sensitivity in determining what is seen
of such Local Group cloud populations. The red symbols in Figures 4 and
5 indicate objects detectable with the relevant LDS or HIPASS
sensitivities, while the black symbols indicate those that remain
undetected. The LDS, in particular, is not sufficiently sensitive to
have detected the majority of CHVCs at distances of more than a few
hundred kpc.
 
The various processes which influence the perceived distributions are
further quantified by de\,Heij et al. (2002b).  About three quarters of
the simulated populations were classified as disrupted due to
ram--pressure or tidal stripping, while 80\% of the remaining objects
were deemed too faint to detect with the LDS (in the north) or HIPASS
(in the south).  Obscuration by Galactic \hi eliminated about half of
the otherwise detectable objects. The total \hi mass involved in the
model populations shown here is $4.3\times10^9$\,M$_\odot$. About 75\%
of this mass had already been consumed by M\,31 and the Galaxy via
cloud disruption, leaving only 25\% still in circulation, although
distributed over some 1200 low--mass objects.

If the de\,Heij et al. models describe the actual distribution of
CHVCs, then the prediction follows that future deeper surveys will
detect large numbers of objects at high negative LSR velocities in the
general vicinity (about $60\times60^\circ$) of M\,31. To make this
prediction more specific, we have imagined the sensitivity afforded by
the current HIPASS survey in the south extended to the entire northern
hemisphere. A high concentration of about 250 faint newly detected
CHVCs is predicted in the Local Group barycenter direction once HIPASS
sensitivity is available, for example from the current Jodrell Bank
HIJASS effort.

One of the most suggestive attributes of the CHVC population in favor
of a Local Group deployment is the modest concentration of objects
which are currently detected in the general direction of M\,31, i.e.
near the direction of barycenter of the Local Group.  These objects
have extreme negative velocities in the GSR reference frame. While this
is a natural consequence of the Local Group models it does not follow
from models simulating a distribution throughout an extended Galactic
halo, nor is it a consequence of obscuration by Galactic H\,{\sc i}.

The choice of a limiting mass of $M_{\rm HI}$=3$\times10^6$\,M$_\odot$
over a linewidth of 35~\kms~was made in the simulation portrayed here
in order to represent what might be possible for a deep \hi survey of
an external galaxy group. In this example, some 188 objects occur which
exceed this mass limit distributed over a region of some 1.5$\times$1.0
Mpc extent. For a limiting mass of $M_{\rm
HI}$=7$\times10^6$\,M$_\odot$ over 35~\kms, the number drops to 50. It
is clear that a very good mass sensitivity will be essential to
detecting such potential CHVC populations in external galaxy
groups. Current searches for such populations, reviewed by Braun \&
Burton (2001), have generally not reached a sensitivity as good as even
$M_{\rm HI}$=$10^7\rm\;M_\odot$ over 35~\kms, so it is no surprise that
such distant CHVCs have not yet been detected.

\section{Properties of individual CHVCs imaged at high resolution}

The all--sky CHVC catalog is based on survey data of modest angular
resolutions: most CHVCs are largely unresolved by the
36\arcmin~resolution of the Dwingeloo telescope, undersampling on a
$0\fdg5$ lattice, although some details in individual CHVCs are
revealed by the 14\arcmin~resolution of the Nyquist--sampled HIPASS
material.  Wakker \& Schwarz (1991) had used the WSRT to image two
CHVCs; subsequently Braun \& Burton (2000) imaged six additional CHVC
fields, and de\,Heij et al.  (2002c) six more.  The angular and
kinematic resolution afforded by the WSRT suit it well in important
regards to detailed studies of the CHVC class of objects.  Diffuse
structures extending over more than about 10\arcmin~are, however, not
adequately imaged by the interferometer unless precautions are taken to
eliminate the short--spacing bowl surrounding regions of bright
extended emission.  Burton et al. (2001) observed several of the WSRT
CHVC fields with the Arecibo telescope, and showed that a combination
of high--resolution filled--aperature and synthesis data is crucial to
determining the intrinsic properties of the CHVCs.  This is especially
important for CHVC targets which are of a size comparable to the field
of view of most synthesis intruments.

\subsection{Nested core/halo morphology}

\begin{figure}
\caption{ Nested core/halo morphology which characterizes those few
CHVCs which have been observed at high resolution by both
interferometric and filled--aperture instruments.  The two left--hand
panels show CHVC\,$186\!+\!19\!-\!114$ as seen with the WSRT and
Arecibo respectively; the right--hand one an overlay for
CHVC\,$230\!+\!61\!+\!165$ with contours showing Arecibo \NH~data;
greyscale for the WSRT data.  The cores are better revealed in
synthesis data than in filled--aperture data because of their small
angular scales; the halos, on the other hand, are better revealed in
filled--aperture data than in synthesis data because of their diffuse
morphology and large angular scales.  Typically only a few tenths of
the total flux observed in a single--dish spectrum is recovered in
synthesis data.}
\end{figure}
 
The CHVCs imaged with the WSRT at resolutions of order $1'$ show a
characterisitic morphology whereby one or more low-dispersion ({\sc
fwhm} line widths in the range 2 to 10 \kms) cores are distributed over
an extent of some tens of arcmin: the larger {\sc fwhm} presented by
the single--dish spectra reflects the differing systemic velocities of
the cores, but is dominated by the intrinsically large velocity
dispersion of the diffuse gas comprising the halo.  Figure 6 shows
combined WSRT and Arecibo data for two CHVCs. The left--hand panel
shows that CHVC\,$186\!+\!19\!-\!114$ comprises some half--dozen cores,
embedded in a common halo.  The unresolved {\sc fwhm} of 20 \kms~shown
by the single--dish spectrum represents the diffuse halo gas as well as
modest differences in the systemic velocities of the narrow--dispersion
cores.  The right--hand panel shows a similar situation for
CHVC\,$230\!+\!61\!+\!165$: several cores, resolved in the WSRT data,
are embedded in a common diffuse envelope seen in the Arecibo
observations, with a {\sc fwhm} of 26 \kms, well represented by a
single Gaussian.

The combined WSRT/Arecibo data thus indicate a morphology of one or
more cores in the Cool Neutral Medium phase, surrounded by a diffuse
halo of Warm Neutral Medium. (Indications of such a core/halo
morphology were found earlier in some extended HVCs, e.g. by Giovanelli
et al. (1981) in detailed \hi~mapping using the Green Bank 300-foot
telescope.) The {\sc fwhm} line width of the CHVC halo gas is typically
25 \kms, consistent with the expected WNM thermal line width of
$10^4$\,K gas.  The {\sc fwhm} of the cores is consistent with CNM
temperatures of order 100 -- 500\,K.  Some of the cores exhibit a
kinematic gradient, consistent with rotation.  The halos evidently do
not share the kinematic gradients of the cores, however, implying that
an external origin of the core gradient, such as tidal shear, for
example, is unlikely.

The Arecibo data have also shown that the halo edge profiles are
exponential, and can be followed to the limiting \NH~sensitivity of
about $2 \times10^{17}$\,cm$^{-2}$.  This exponential form implies that
measurements of apparent angular sizes and total flux densities of CHVC
halos will depend on the resolution as well as on the sensitivity of
the data; larger surface covering factors will be returned by deeper
observations (see Lockman et al. 2002). The exponential edges also
suggest that the outer envelopes of CHVCs are not tidally truncated,
consistent with the objects residing at substantial distances from the
Milky Way. The sources studied by Burton et al. (2001) indicate a
characteristic central halo column density of $4.1 \pm3.2\times
10^{19}$\,cm$^{-2}$, and a characteristic exponential scalelength of
$420\pm90$\,arcsec.  For plausible values of the thermal pressure at
the core/halo interface, these edge profiles support distance estimates
which range between 150 and 850 kpc. Similar distances are indicated by
simply equating the observed neutral scalelengths with those predicted
to occur at the edges of low mass galactic disks of about 1~kpc
(Corbelli and Salpeter 1993) over a wide range of phsyical conditions.
   
\subsection{Rotation of cores}

\begin{figure}
\caption{ {\it left panel:} Overlay of WSRT and Arecibo $N_{\rm HI}$
data for CHVC\,$204\!+\!30\!+\!075$, showing several cores embedded in
a common diffuse halo.  Seen unresolved in the $36'$ beam of the 25--m
Dwingeloo telescope, this source has a velocity {\sc fwhm} of 34 \kms.
{\it upper right:} Intensity--weighted line--of--sight velocity,
derived from WSRT data at an angular resolution of 1\arcmin.  Contours
of $V_{\rm LSR}$ show systematic kinematic gradients across the two
principal components of the CHVC, consistent with rotation; contours
are drawn in steps of 5 \kms~from 40 to 85 \kms.  ~{\it lower right:}
Rotation velocities resulting from the tilted--ring model applied to
the kinematic gradient of the lower--declination component of
CHVC\,$204\!+\!30\!+\!075$.  The best--fit position, inclination, and
systemic velocity are indicated. The solid line shows the rotation
curve corresponding to a Navarro et al. (1997) CDM halo of the
indicated mass.}
\end{figure}

Under higher angular resolution than afforded by the LDS or HIPASS, the
characteristic linewidth of individual CHVC components narrows as
objects are resolved into several principal units, moving relative to
each other.  If the objects with multiple cores are to be stable,
distances of order several hundred kpc are required.  Some of the
compact cores imaged owe their large total width seen in single--dish
data to velocity gradients.  Several of the CHVC cores resolved with
the WSRT exhibit kinematic gradients consistent with rotation.  Figure
7 shows that the resolved WSRT image of CHVC\,$204\!+\!30\!+\!075$ has
two principal components, each of which is elongated and shows,
furthermore, a systematic velocity gradient along the major axes.  The
velocity gradients exhibited by the two principal components of
CHVC\,$204\!+\!30\!+\!075$ can be modelled in terms of circular
rotation in a flattened disk system.  Figure 7 shows the results of
fitting a standard tilted--ring model to one of the cores.  The fit
shows velocity rising slowly but continuously with radius to an
amplitude of some 20 \kms, and then flattening to a constant value
beyond about 500 or 600 arcsec.  Estimates of the contained dynamical
mass follow from the rotation curve if the distance is assumed, and the
total gas mass follows from the integrated \hi flux.  At an assumed
distance of 0.7 Mpc, the two principal clumps of
CHVC\,$204\!+\!30\!+\!075$ have $M_{\rm dyn}=10^{8.1}$ and $10^{8.3}$
M$_\odot$, and gas masses (including \hi and helium of 40\% by mass) of
$10^{6.5}$ and $10^{6.9}$ M$_\odot$, respectively for the upper and
lower concentrations shown in Fig. 7. The dark--to--visible mass ratios
for these concentrations are 36 and 29, respectively.  The shape of the
modelled rotation curves for both of the CHVC\,$204\!+\!30\!+\!075$
components resembles that of the standard cold--dark--matter halo as
presented by Navarro et al.  (1997).  At the assumed distance of 0.7
Mpc, the Navarro et al. halos fit to the two components have masses of
$10^{7.8}$ M$_\odot$ (within 9.3 kpc) and $10^{8.2}$ M$_\odot$ (within
12.6 kpc), respectively.

\subsection{A very cold core}

\begin{figure}
\caption{ {\it upper panel:} WSRT image of CHVC\,$125+41-207$
displaying the $N_{\rm HI}$ distribution at $28''$ angular resolution.
The Seyfert galaxy Mrk\,205 lies on a line of sight which penetrates
the halo of the CHVC, reconstructed using the composite WSRT and LDS
single--dish data.  {\it lower left:} \hi spectrum observed in the
direction (indicated in the upper panel) of one of the bright emission
knots.  The spectrum is unresolved at a channel separation of 2 \kms,
indicating a core temperature of less than 85\,K and quiescent
turbulence.  {\it lower right:} Equilibrium temperature curves for \hi
in an intergalactic environment (with a metallicity of 0.1 solar and a
dust--to--gas mass ratio of 0.1 times that characteristic of the solar
neighborhood), shown for two values of the neutral shielding column
density, namely $10^{19}$ cm$^{-2}$ (solid line) and $10^{20}$
cm$^{-2}$ (dashed line). The 85\,K kinetic temperature tightly
constrains the volume density; a distance of order 500 kpc then follows
from the measured column density and angular size. }
\end{figure}

CHVC\,$125\!+\!41\!-\!207$ is representative of the compact objects in
several regards. The WSRT image of CHVC\,$125\!+\!41\!-\!207$ observed
by Braun \& Burton (2000) and shown in Fig. 8 reveals several cool,
quiescent cores embedded in a diffuse, warmer halo. The spectrum
plotted in the lower left of the figure was observed towards the
brightest of these cores: it has a line width unresolved by the 2
\kms~resolution of the WSRT imaging. The velocity channels adjacent to
the line peak have intensities below 20\% of the maximum value.  Such a
width is one of the narrowest measured in \hi emission --- \hi line
widths are almost always dominated by line blending and mass motions --
and constrains both the kinetic temperature and the amount of
turbulence within the core.  An upper limit to the thermal--broadening
{\sc fwhm} of 2 \kms~corresponds to an upper limit to the kinetic
temperature of 85 K.  As it happens, the physical situation is yet more
tightly constrained, because the brightness temperature in this core
was observed to be 75 K; thus a lower limit to the opacity follows from
$T_{\rm b} = T_{\rm k}(1 - e^{-1})$, yielding $\tau \geq 2$. Any
broadening which might be due to macroscopic turbulence is less than 1
\kms. (Such quiescence would be most unusual in a Milky Way
environment.)
 
A fortunate, albeit fortuitous, property of this CHVC is that it lies
near the direction of the Seyfert galaxy Mrk\,205. (In general, the
small angular size of CHVCs, and the amount of substructure being
revealed at high resolution, will render it difficult to find suitable
background sources to support metallicity measurements.)  Bowen \&
Blades (1993) measured Mg\,{\sc ii} absorption towards Mrk\,205,
finding a metallicity substantially subsolar.  Wolfire et al. (1995)
show that a cool \hi phase is stable under extragalactic conditions if
a sufficient column of shielding gas is present and if the thermal
pressure is high.  Calculations of equilibrium conditions which would
pertain in the Local Group environment characterized by a metallicity
of 10\% of the solar value and a dust--to--gas ratio of 10\% of that in
the solar neighborhood, were communicated to us by Wolfire, Sternberg,
Hollenbach, and McKee.  The tightly--constrained temperature found for
CHVC\,$125\!+\!41\!-\!207$ then allows an estimate of the distance to
this object.  Figure 8 shows two bracketing values of the shielding
column density, namely $10^{19}$~cm$^{-2}$ and $10^{20}$~cm$^{-2}$.
The equilibrium volume densities corresponding to the observed value of
$T_{\rm k}=85$~K lie in the range 0.65 to 3.5~cm$^{-3}$. Provided with
this range of volume densities, and having measured both the column
depth of the cool core and its angular size, the distance to
CHVC\,$125\!+\!41\!-\!207$ follows directly from $D=N_{\rm HI}/(n_{\rm
HI}/\theta)$, yielding a value in the range 210 to 1100 kpc.

\section{Concluding remarks}

The concept of a distinct class of compact, isolated high--velocity
clouds has been objectively developed by application of a search
algorithm to the LDS and HIPASS datasets, resulting in an all--sky
catalog of CHVCs, high--contrast \hi features which are at best only
marginally resolved with half--degree angular resolution and which are
sharply bounded, such that emission at column densities above a few
times 10$^{18}$ cm$^{-2}$ is unconfused by extended emission in the
source environment. We identify five principal observables from the
all--sky catalog: the spatial and kinematic deployments, and the number
distributions of angular sizes, \NH, and linewidth. We show that
agreement of models with the data must be judged by extracting these
same obervables from the simulations in a manner consistent with the
sensitivities of the observations, and explicitly taking into account
the limitations imposed by obscuration due to foreground \hi in the
Milky Way.  General considerations of these observables show
compatibility with characteristic distance scales of hundreds of kpc.

Since the inception of high--velocity cloud research, the possibility
of an extragalactic deployment of these clouds has been critically
considered as a possibility; among others, the discussions by Oort
(1966, 1970, 1981), Verschuur (1969, 1975), Giovanelli (1981), Bajaja,
Morras, \& P\"oppel (1987), Wakker \& van Woerden (1997), Blitz et
al. (1999), and BB99 are particularly relevant here. Our simulations of
a Local Group population of CHVCs have accounted explicitly for the
obscuration resulting from the foreground Galactic \hi and the
observational attributes of the existing survey material. 
We have shown that models in which the CHVCs are the \hi
counterparts of dark--matter halos evolving in the Local Group can
provide a good match to the observables, if the simulations are
``observed'' in accordance with the LDS and HIPASS parameters.

High--resolution imaging confirms the nested core/halo geometry
expected if the CNM is to be stable in the presence of an ionizing
radiation field of the sort expected in the Local Group
environment. The cores contribute typically about 40\% of the \hi flux,
while covering about 15\% of the surface of the CHVC.  The imaging
also allows supports two independent distance estimates.  A distance
for CHVC\,$125\!+\!41\!-\!207$ follows from assuming rough spherical
symmetry and equating the well--constrained volume and column densities
of the compact cores. Another distance constraint (coupled with a
dark--to--visible mass ratio) follows from consideration of the
stability of CHVCs having multiple cores in a common envelope but
having large relative velocities.

The available evidence suggests that CHVCs have characteristic sizes of
about 2 kpc, \hi masses in the range $10^{5.5}$ to $10^7$ M$_\odot$,
and are typically seen at distances of hundreds of kpc.  If at such
distances, the failure to detect stars would imply that CHVCs are very
primitive proto--galactic objects dominated by dark--matter halos,
plausibly the the missing Local Group satellite systems predicted by
Klypin et al. (1999) and Moore et al. (1999).

{}
 
\end{document}